\documentclass[pdflatex,sn-mathphys-num]{sn-jnl}

\usepackage{graphicx}%
\usepackage{multirow}%
\usepackage{amsmath,amssymb,amsfonts}%
\usepackage{amsthm}%
\usepackage[title]{appendix}%
\usepackage{xcolor}%
\usepackage{textcomp}%
\usepackage{manyfoot}%
\usepackage{booktabs}%
\usepackage{algorithm}%
\usepackage{algorithmicx}%
\usepackage{algpseudocode}%
\usepackage{listings}%

\usepackage{comment}

\begin{document}

\title[Article Title]{Effect of Interatomic Potential Choice on Fracture Modes of Graphene with Parallel Cracks}


\author[1]{\fnm{Suyeong} \sur{Jin}}\email{suyeong@pknu.ac.kr}
\author[2]{\fnm{Jung-Wuk} \sur{Hong}}\email{j.hong@kaist.ac.kr; jwhong@alum.mit.edu; jungwukh@gmail.com}
\author*[3]{\fnm{Alexandre F.} \sur{Fonseca}}\email{afonseca@ifi.unicamp.br}

\affil[1]{\orgdiv{Department of Mechanical Engineering}, \orgname{Pukyong National University}, \orgaddress{\street{45 Yongso-ro, Nam-gu}, \city{Busan}, \postcode{48513}, \country{Republic of Korea}}}
\affil[2]{\orgdiv{Department of Civil and Environmental Engineering}, \orgname{Korea Advanced Institute of Science and Technology}, \orgaddress{\street{291 Daehak-ro, Yuseong-gu}, \city{Daejeon}, \postcode{34141}, \country{Republic of Korea}}}
\affil*[3]{\orgdiv{Departamento de F\'{i}sica Aplicada}, \orgname{Instituto de F\'{i}sica Gleb Wataghin, Universidade Estadual de Campinas}, \orgaddress{\city{Campinas}, \state{SP},\postcode{13083-859}, \country{Brazil}}}

\abstract{Defect engineering via parallel cracks has been proposed as
  a route to tailor the fracture response of graphene. However,
  atomistic fracture predictions can be strongly sensitive to the
  interatomic potential. Here, we quantify the effect of potential
  choice by revisiting H-passivated graphene containing two parallel
  cracks separated by a gap $W_\text{gap}$ loaded in tension along the
  armchair (AC) and zigzag (ZZ) directions. Molecular dynamics
  simulations using the AIREBO potential under the same geometry and
  loading protocol previously studied with ReaxFF, are employed, so
  enabling a direct comparison. Stress–strain responses, Young’s
  modulus, an effective mode-I stress intensity factor, and energy
  absorption are evaluated as functions of $W_\text{gap}$. Compared
  with ReaxFF, AIREBO predicts lower peak stresses and earlier
  catastrophic softening, leading to reduced post-peak deformation
  capacity and energy absorption. Ductility and energy absorption are
  shown to be highly potential-dependent, underscoring the need for
  careful potential selection in defect-engineered graphene fracture
  simulations.}


\maketitle


\section{Introduction} Graphene has garnered sustained attention as a
structural and functional material due to its exceptional mechanical,
electrical, and thermal properties, rendering it a promising candidate
for applications ranging from flexible electronics to nano-reinforced
composites~\cite{Novoselov2004Science}. Despite the passage of more
than two decades since its initial extraction, graphene continues to
impress the scientific community, as demonstrated by recent
advancements in biomedicine~\cite{Islam2024SensorsI}, thermoelectric
materials~\cite{Xue2025RSCAdv}, nanocomposites for high-performance
energy storage~\cite{Raj2025Ionics}, and photovoltaic
devices~\cite{Jain2024ACSOmega}, among others~\cite{Science2024}.

In practical implementations, graphene sheets inevitably contain
defects introduced during synthesis and processing~\cite{Liang2011}.
Recent studies have investigated graphene structures with artificially
introduced defects, such as notches, holes, and parallel cracks.
These defects can be exploited to tailor the mechanical response of
graphene, including stiffness, strength, ductility, and energy
dissipation~\cite{Zhang2014, Zhang2012, Meng2015, DewaCMS2018}. As a
result, defect-engineered graphene has emerged as an important design
paradigm for achieving property enhancement rather than degradation.

Recently, some of the present authors reported an interesting
mechanical behavior in graphene~\cite{suyeong2026IJMS}. They
demonstrated the possibility of controlling a brittle-to-ductile
transition in graphene with pre-existing cracks by tailoring the
crack-to-crack distance, $W_{\text{gap}}$. Using classical molecular
dynamics (MD) simulations, the reactive force field (ReaxFF)
potential~\cite{vanDuin2021} was used to examine the process of crack
coalescence under tensile loading. The work showed that above a
certain distance, $W_{\text{gap}}$, graphene exhibited a more
ductile-like fracture response with increased energy absorption.  This
phenomenon is noteworthy, as prior studies have reported that graphene
typically exhibits brittle fracture behavior at any temperature below
its melting point~\cite{Antonio2022CARBONTRENDS}. This result
indicates that $W_{\text{gap}}$ can be tailored to manage and design a
brittle-to-ductile behavior in graphene.

The above finding of a brittle-to-ductile behavior is compelling, but
it still lacks experimental confirmation. Moreover, at the atomistic
scale, fracture predictions are known to be sensitive to the
interatomic potential used because crack-tip mechanics involve highly
nonlinear bond stretching, angular interactions, and bond-order
transitions~\cite{suyeong2026IJMS}.  The simulation of intrinsic
strength and failure of graphene can be governed by nonlinear
elasticity, where higher-order stiffness terms become dominant under
extreme stress concentrations at crack or indenter
tips~\cite{Changgu2008}, and large deformations can induce anomalous
behavior such as a negative Poisson’s ratio via coupled bond-angle and
bond-length evolution, accompanied by anisotropy and strain
softening~\cite{Pelliciari2021}.  Therefore, the choice of interatomic
potential can significantly affect the predicted fracture initiation
and crack propagation paths, as differences in bond-order evolution
and higher-order stiffness representation govern strain softening and
bond dissociation within the fracture process zone.

The purpose of this study is to evaluate the persistence of the
previously reported $W_{\text{gap}}$-mediated brittle-to-ductile
transition when an alternative MD potential is employed for the same
simulations.  Among the interatomic potentials commonly used for
carbon-based materials, the reactive force field (ReaxFF) and adaptive
intermolecular reactive empirical bond order (AIREBO) potentials have
been widely employed to investigate fracture and failure in
graphene~\cite{Zacharias2022Frontier,Bu2022ASurfSci,Zhang2012,Zhang2014,Yin2015NL,FonsecaGalvao2019Carbon}. ReaxFF
permits continuous bond formation and dissociation via dynamic charge
equilibration, making it particularly suitable for chemically reactive
environments and complex bond rearrangements~\cite{vanDuin2021}.  In
contrast, AIREBO describes bond-order–dependent interactions based on
the neighborhood of the pair of atoms forming the bond, augmented by
torsional and long-range Lennard--Jones terms, enabling simulations of
bond breaking and large deformations in covalently bonded carbon
systems~\cite{brenner2002,stuart2000}. AIREBO has been applied
extensively to study different physical properties of
graphene~\cite{Rassin2010,NeekAmal2011PRB,FonsecaMuniz2015JPCC,Fonseca2021PRB},
including crack formation and
propagation~\cite{Zhang2012,Zhang2014,Yin2015NL,FonsecaGalvao2019Carbon}.
However, a systematic comparison of fracture modes, particularly in
defect-engineered graphene geometries where crack interaction and
coalescence dominate failure behavior, remains limited.

Motivated by this gap, we re-examined the $W_{\text{gap}}$-dependent
fracture behavior using AIREBO under identical geometric and loading
conditions.  The fracture modes of graphene containing parallel cracks
using AIREBO were investigated and compared with those obtained using
ReaxFF~\cite{suyeong2026IJMS} under identical geometric and loading
conditions. We focus on differences in crack initiation, propagation,
and coalescence, and discuss implications of potential choice for
atomistic fracture modeling. Through the comparison, this study will
demonstrate the effect of interatomic potential choice and provide a
methodological basis for atomic-level fracture simulations.


\section{\label{sec:model} Model Description and Computational Methods}
In this section, the structures and the computational methodologies
are described. In subsection~\ref{graphene}, the geometry of the
cracks is presented. Then, in subsection~\ref{protocols}, the
classical potential and simulation protocols are described. The
simulations are performed in LAMMPS~\cite{lammps2022}.

\subsection{Graphene structure}\label{graphene}

\begin{figure}[]
    \centering
    \includegraphics[width=0.95\linewidth]{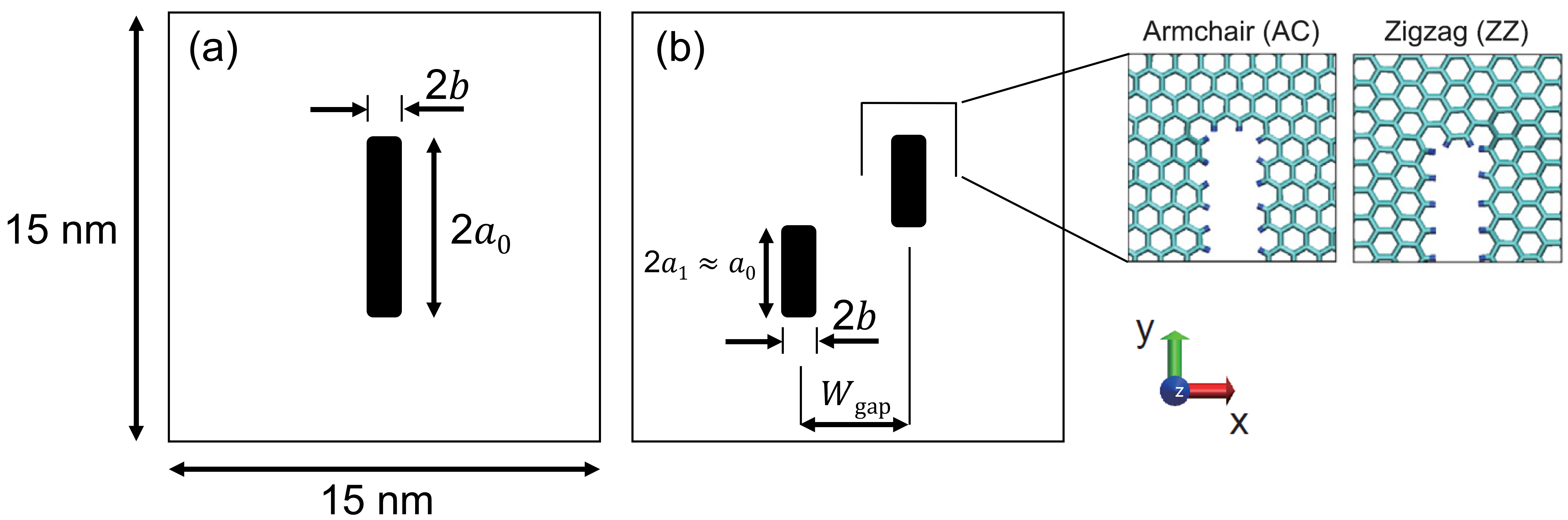}
    \caption{\label{fig:geometry}Geometry of graphene with (a) a
      single crack of length 2$a_0$ and (b) parallel cracks, each of
      length 2$a_1$, where $2a_1\approx a_0$, separated by a gap
      $W_\text{gap}$. The local atomic structure around the cracks is
      shown in a magnified view, with carbon and hydrogen atoms
      colored cyan and blue, respectively.}
\end{figure}

A scheme illustrating a graphene sheet with a single crack is shown in
Fig.~\ref{fig:geometry}(a), with a crack length of $2a_0$ and a width
of $2b$. Graphene samples with parallel cracks are also prepared, with
the structure shown in Fig.~\ref{fig:geometry}(b), where the cracks
are separated by a gap $W_\text{gap}$, and each crack has a length of
$2a_1\approx a_0$. When $W_\text{gap}$ becomes zero, the cracks are
merged to form a single crack with a length of $2a_0$.

\subsection{Simulation Protocols}\label{protocols}

All molecular dynamics simulations are performed in
LAMMPS~\cite{lammps2022} using metal units and the AIREBO potential,
with the same H-passivated graphene configurations studied in
Ref.~\cite{suyeong2026IJMS}. The structure is relaxed by energy
minimization with tolerance criteria set to 10$^{-9}$ (dimensionless)
for energy and 10$^{-9}$ eV/\AA~for force.  After energy minimization,
a NVE time integration with the Langevin thermostat is applied to the
unconstrained atoms to maintain and equilibrate the systems at 300~K.
Then, for the tensile simulation, the left edge is fixed, and the
Dirichlet boundary condition (BC) is applied to the right edge at a
strain rate of $10^8~\text{s}^{-1}$ ($10^{-7}~\text{fs}^{-1}$).  The
frames with the snapshots of the structures during the tensile strain
simulations are monitored in order to finish the simulation as soon as
the structure is broken in two separate parts.  Stress–strain curves
are obtained by computing the per-atom virial stress using
\textit{stress/atom} LAMMPS commands for the atoms within the free
region and summing the stress components over this group.  The
continuum stress (in GPa) is then evaluated by dividing the summed
virial stress by an effective volume defined, $V_\text{eff}=tL_xL_y$,
where $t=3.34$~\AA~is the assumed graphene thickness and $L_x$ and
$L_y$ are the effective length and width of the unconstrained graphene
region.  For data analysis, Young’s modulus is obtained by linear
regression of the stress–strain curve over the strain range from zero
to 0.01.  The critical effective stress intensity factor in mode-I,
$K^{\mathrm{eff}}_{\mathrm{IC}}$, is evaluated following the
definition in Ref.~\cite{suyeong2026IJMS}, with $a^{\mathrm{eff}}=a_0$
for $W_\text{gap}=0$ and $a^{\mathrm{eff}}=2a_1$ for
$W_\text{gap}\neq0$.  The energy absorption is calculated as the area
under the stress–strain curve over the strain range from zero to 0.08.
The von Mises~\cite{vonMises1913} stresses are calculated as explained
in Ref.~\cite{suyeong2026IJMS}.

\section{Results}

\begin{figure}[tbp]
    \centering
    \includegraphics[width=\linewidth]{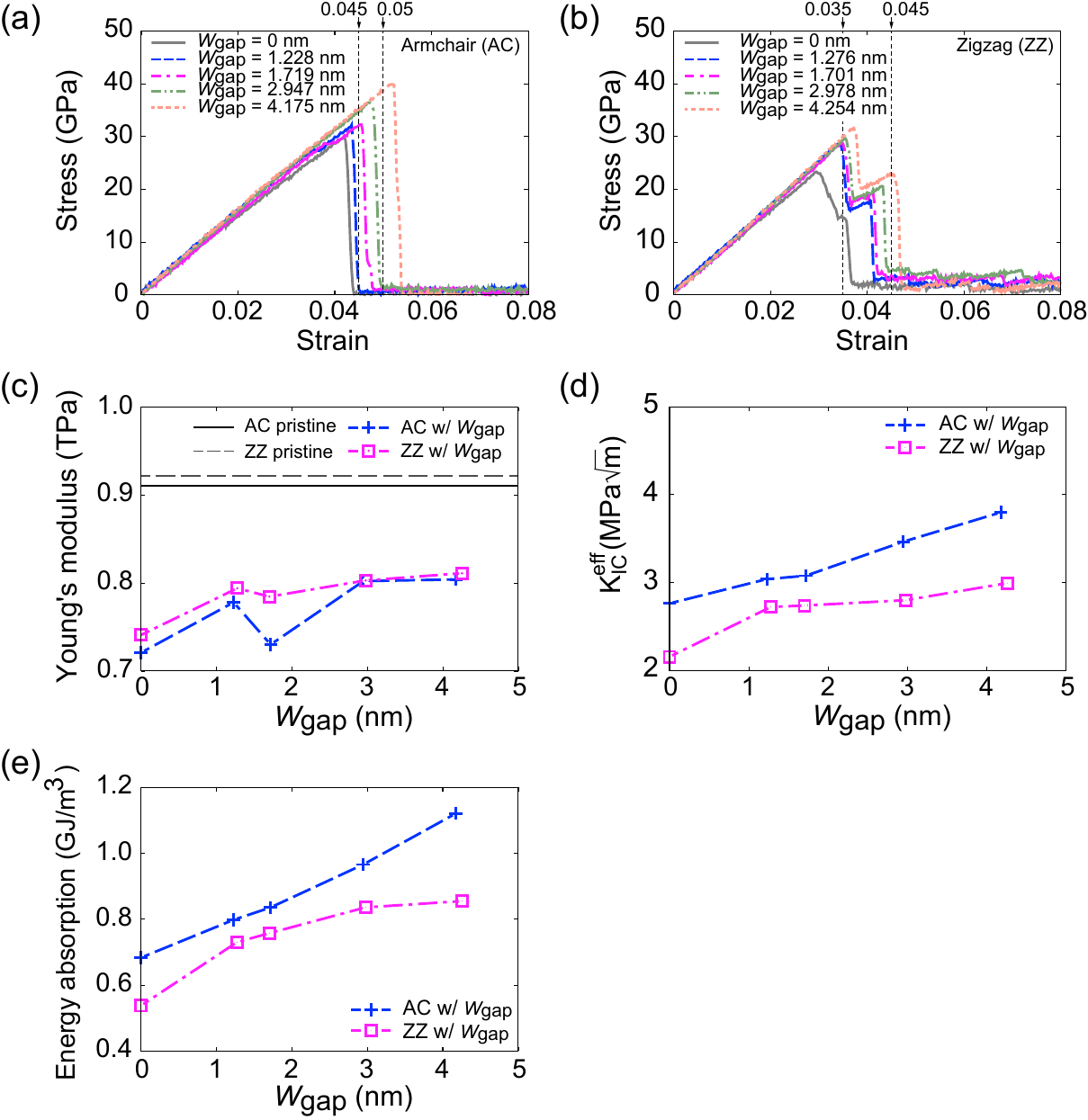}
    \caption{\label{fig:ss-curve-main}Results of armchair (AC) and
      zigzag (ZZ) structures. (a, b) Stress versus strain curves for
      armchair and zigzag, respectively, with varying $W_\text{gap}$,
      at a strain rate of $10^8$~s$^{-1}$. (c) Young's modulus versus
      $W_\text{gap}$ for both AC and ZZ structures. (d) Effective
      stress intensity factor versus $W_\text{gap}$, corresponding to
      the stress-strain curves shown in (a) and (b). (e) Energy
      absorption under the stress-strain curve. The vertical dashed
      lines at strain values are references for the next figures.}
\end{figure}

Fig.~\ref{fig:ss-curve-main} presents the tensile simulation results
for graphene loaded along the armchair (AC) and zigzag (ZZ)
directions, obtained using the AIREBO potential and computational
protocols described in section~\ref{sec:model}.
Figs.~\ref{fig:ss-curve-main}(a) and (b) summarize the stress-strain
responses of graphene with armchair and zigzag chiralities,
respectively, containing parallel cracks separated by a distance
$W_\text{gap}$.  The stress–strain curves exhibit an almost linear
elastic regime followed by a sharp stress drop at a critical strain,
indicating abrupt loss of load-carrying capacity once fracture
initiates.  In particular, for armchair cracks, the peak stress
increases as $W_\text{gap}$ increases, and the critical strain at peak
also shifts to larger values.  The peak stress rises from
approximately 30~GPa at $W_\text{gap}=0$ to 40~GPa at the largest
separation, while the peak strain increases from roughly $0.04$ to
$0.05$.  For the zigzag configuration, the peak stresses are lower,
approximately 23-31~GPa, and occur at smaller strains, approximately
at 0.03-0.04, compared with the AC case. Moreover, the zigzag case
exhibits a two-step softening behavior, characterized by an initial
partial stress drop followed by a second drop. This two-step softening
behavior is discussed in detail in the next section.

The AIREBO-based Young’s modulus of pristine graphene is approximately
0.91 TPa for armchair and 0.92 TPa for zigzag. Introducing parallel
cracks reduces the effective modulus, particularly for the
merged-crack case $\left(W_\text{gap}=0\right)$, where the modulus
drops to about $0.72–0.74$~TPa. With increasing $W_\text{gap}$, the
modulus partially recovers and approaches 0.80–0.81 TPa at the largest
separations for both orientations.

Consistent with the peak trend, the effective fracture toughness
$K^\text{eff}_\text{IC}$ increases with $W_\text{gap}$, and AC remains
higher than ZZ across all cases. $K^\text{eff}_\text{IC}$ increases
from approximately 2.7 to 3.8~MPa$\sqrt{\text{m}}$ for AC and from 2.1
to 3.0 MPa$\sqrt{\text{m}}$ for ZZ as $W_\text{gap}$ increases.  These
values are within those obtained from other
computational~\cite{suyeong2026IJMS,Le2016} and
experimental~\cite{HwaSR2014} data.

Overall, within the AIREBO framework, increasing $W_\text{gap}$
enhances the tensile performance, such as higher peak,
$K^\text{eff}_\text{IC}$, and energy absorption. Also, the AC-oriented
cracks exhibit higher resistance compared to ZZ-oriented cracks. The
vertical dashed lines in Fig.~\ref{fig:ss-curve-main}(a,b) indicate
the two selected strain levels, 0.045 and 0.05 for AC and 0.035 and
0.045 for ZZ, used to extract the fracture snapshots presented in
Fig.~\ref{fig:fracture-snapshots}.

\section{Discussion}

\begin{figure}[tbp]
    \centering
    \includegraphics[width=\linewidth]{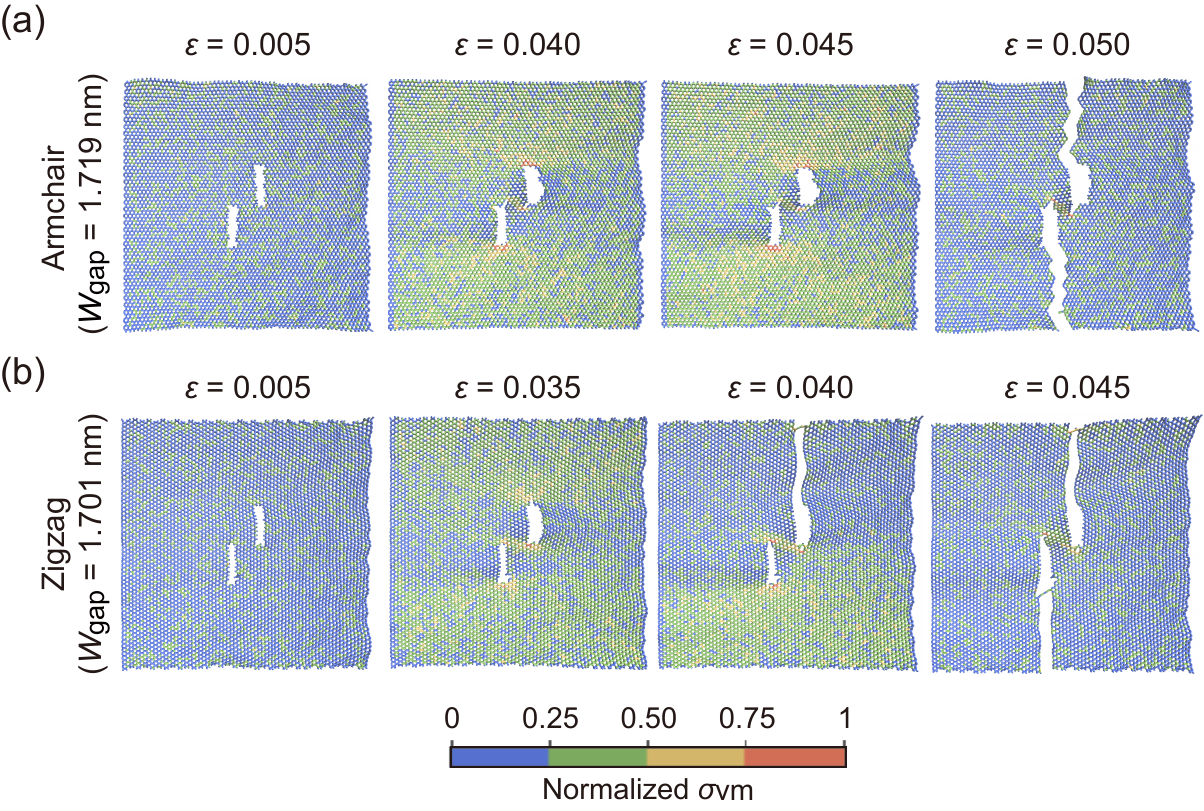}
    \caption{\label{fig:fracture-snapshots}Snapshots of fracture
      evolution at selected strain levels $\varepsilon$, with atoms
      colored by normalized von Mises stress $\sigma_\text{vm}$: (a)
      armchair configuration with $W_\text{gap}=1.719$~nm at
      $\varepsilon=0.005, 0.040, 0.045,$ and $0.050$; (b) zigzag
      configuration with $W_\text{gap}=1.701$~nm at
      $\varepsilon=0.005, 0.035, 0.040,$ and $0.045$.}
    \end{figure}
In the AIREBO results, the distinct post-peak softening behaviors
between AC and ZZ can be explained by the fracture sequence observed
in Fig.~\ref{fig:fracture-snapshots}. In the armchair case
($W_\text{gap}=1.719$~nm), both outer cracks become unstable nearly
simultaneously, producing a single abrupt stress drop at
$\varepsilon=0.045-0.050$, reflecting catastrophic failure with
minimal intermediate load redistribution.  In contrast, the ZZ case
($W_\text{gap}=1.701$~nm) exhibits a more staged fracture
evolution. One of the initial cracks begins to propagate outward
approximately at $\varepsilon = 0.040$, followed by complete
separation at $\varepsilon = 0.045$. This stepwise progression is
consistent with the two-step stress drop observed in the corresponding
stress-strain curve, indicating sequential fracture events: the first
drop marks the onset of outward propagation from one outer crack tip,
while the second drop corresponds to fracture of the remaining crack
propagating outwardly and the ensuing loss of load-carrying capacity.

Having established that the one-step (AC) versus two-step (ZZ)
stress-drop behavior is governed by the fracture sequence of the outer
cracks (Fig.~\ref{fig:fracture-snapshots}), we next assess how
sensitive these trends are to the choice of interatomic
potential. Specifically, the AIREBO-based responses are compared with
our previously reported ReaxFF results obtained under the same
geometric and loading conditions~\cite{suyeong2026IJMS}.

Comparing with the AIREBO-based results, ReaxFF, reported in
Ref.~\cite{suyeong2026IJMS}, predicts a substantially higher
load-carrying capacity and a more extended post-peak response than
AIREBO for both armchair (AC) and zigzag (ZZ) crack configurations.
In the stress-strain curve obtained using ReaxFF, the peak stress
reaches roughly 65–90 GPa around $\varepsilon\approx0.075$, followed
by a long, gradually decaying tail that persists up to
$\varepsilon\approx0.3$, indicating sustained load transfer after peak
and a more ductile-like macroscopic response.  In contrast, AIREBO
predicts earlier catastrophic softening with a sharp stress drop at
significantly lower strains $\varepsilon\approx0.03-0.05$ and
comparatively limited post-peak load-bearing (stress rapidly decreases
to near-zero shortly after peak), consistent with a more brittle-like
response at the continuum level. Peak stresses are also smaller,
approximately $23–40$~GPa.  About energy absorption, ReaxFF
consistently yields substantially larger energy absorption than AIREBO
across $W_\text{gap}$. The difference arises from the higher stress
level before peak and the sustained post-peak load transfer predicted
by ReaxFF, which further increases the area under the stress–strain
curve.  In contrast, under the AIREBO potential, the peak stress is
lower and the post-peak response exhibits an abrupt stress drop,
reducing the post-peak contribution and leading to lower overall
energy absorption.

Despite the differences in the absolute values of the quantities shown
in Fig.~\ref{fig:ss-curve-main}, both AIREBO and ReaxFF exhibited
certain similar trends, suggesting a few results that are MD-potential
independent. As observed in the preceding
study~\cite{suyeong2026IJMS}, the peak stresses and effective fracture
toughness increase with $W_\text{gap}$ and are greater for AC
structures than for ZZ structures. A comparison of
Fig.~\ref{fig:ss-curve-main}(e) with Figure~5(e) of
Ref.~\cite{suyeong2026IJMS} reveals similar trends.
The energy absorption of AC structures consistently exceeds that of ZZ
structures and a similar convergence behavior of the energy absorption
of ZZ structures is observed for large $W_\text{gap}$. This result is
of particular interest and should be tested in future experiments.

\begin{figure}[tbp]
    \centering
    \includegraphics[width=\linewidth]{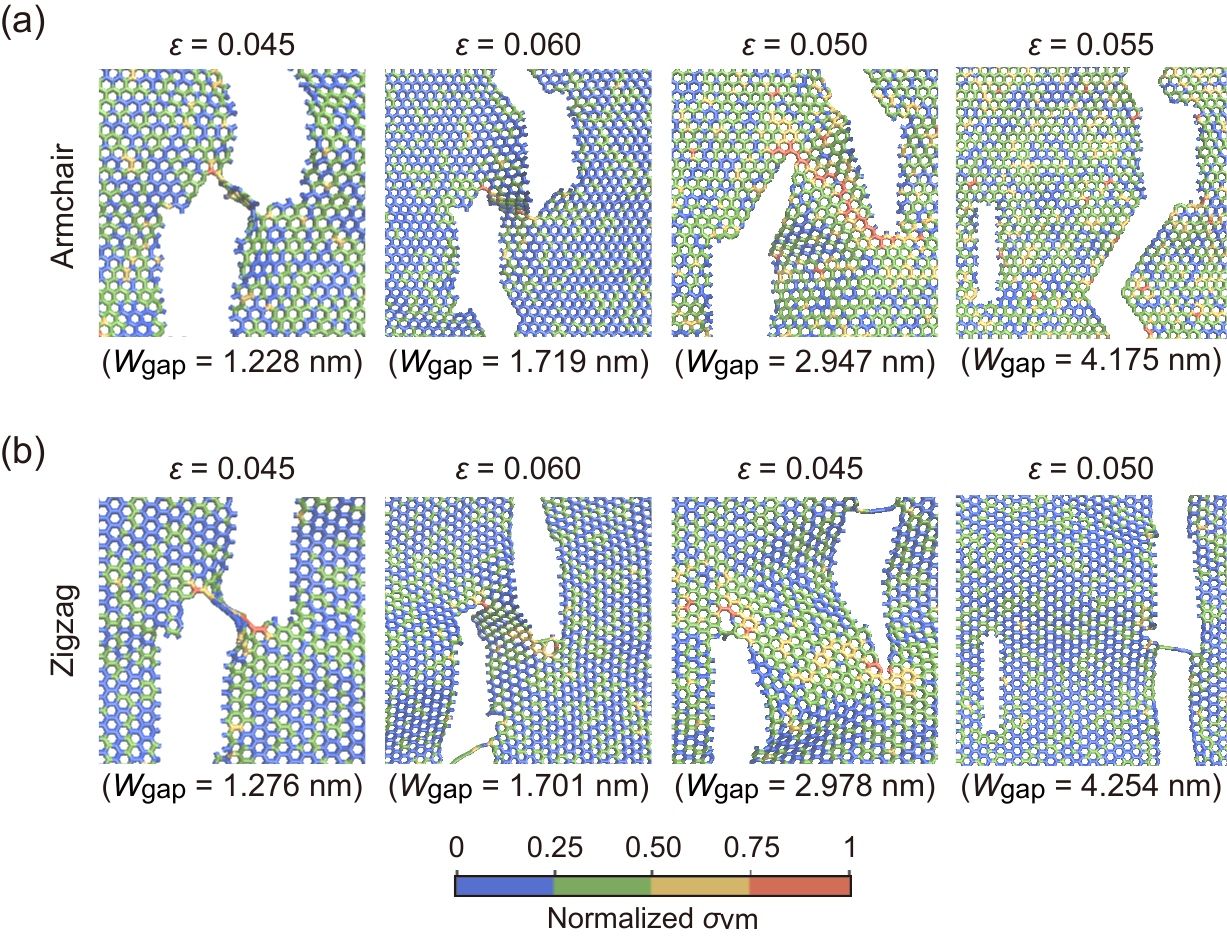}
    \caption{\label{fig:fractureMode} Lever-like structures formed in
      the region between the inner crack tips for (a) armchair and (b)
      zigzag configurations, except the largest-$W_\text{gap}$ case
      showing one crack propagation. Snapshots are shown for varying
      $W_\text{gap}$ at selected strain levels $\varepsilon$. The
      color contours denote the normalized von Mises stress,
      $\sigma_\text{vm}$, where $\sigma_\text{vm}$ is normalized by
      the maximum value in each corresponding case.}
\end{figure}

A final noteworthy observation, albeit qualitative, concerns the
fracture mode in certain structures.  In our previous
study~\cite{suyeong2026IJMS}, a ductile-like fracture response emerged
when the initial crack spacing $W_\text{gap}$ exceeded a certain
value. In that regime, the two cracks did not coalesce; the ligament
between the inner crack tips remained bonded until complete
separation. Those cases were classified as exhibiting a “lever”-like
type of fracture~\cite{suyeong2026IJMS}.  In the present study,
however, the AIREBO simulations do not show prior coalescence of the
inner crack tips, even at the smallest non-zero $W_\text{gap}$. In
other words, inner-tip coalescence does not precede outward crack
propagation. As illustrated in Fig.~\ref{fig:fractureMode}, a
lever-like ligament forms between the two cracks for most
configurations with $W_\text{gap}$ below a critical level. When
$W_\text{gap}$ becomes sufficiently large, the specimen instead fails
by propagation of only one crack to rupture, a behavior that was not
observed in the ReaxFF results. This feature of forming a
``lever''-like structure between the cracks, that are common in the
results of the simulations with both AIREBO and ReaxFF, is also worthy
of experimental confirmation, as graphene structures showing such a
``lever''-like fracture will keep, at least in part, the eletrical
conductivity between their two extremes and, consequently, could be
useful in sensors.

\section{\label{sec:Conclusions} Conclusions}
This study revisited the $W_\text{gap}$-dependent fracture behavior of
H-passivated graphene with parallel initial cracks using the AIREBO
potential and assessed the sensitivity of the predicted response by
comparison with our previously published ReaxFF
results~\cite{suyeong2026IJMS} under the same geometric and loading
conditions.  The interatomic potential AIREBO reproduces a clear
$W_\text{gap}$-dependent strengthening trend. Also, the post-peak
softening pattern depends on crack orientation and fracture
sequence. The predicted values of the macroscopic energy absorption is
highly potential-sensitive based on the comparison, but the trend with
$W_\text{gap}$ is similar to that obtained with
ReaxFF~\cite{suyeong2026IJMS}.  Overall, the beneficial effect of
increasing $W_\text{gap}$ is qualitatively robust. However, the peak
stress, post-peak deformation capacity, and energy absorption -- and
the resulting assessment of brittle-to-ductile behavior -- is shown
here to strongly depend on the interatomic potential, underscoring the
need for careful potential selection and interpretation in atomistic
fracture simulations of defect-engineered graphene.  A key limitation
of this study is the lack of direct validation against experimental
fracture data; accordingly, the present results should be interpreted
as potential-dependent trends rather than as evidence for the
superiority of any single interatomic potential for defect-engineered
graphene.

\section*{Acknowledgments} 
This work used resources of the John David Rogers Computing Center
(CCJDR) in the Gleb Wataghin Institute of Physics, University of
Campinas. Special computational resources were provided by the
Coaraci Supercomputer (S\~{a}o Paulo Research Foundation (FAPESP)
grant \#2019/17874-0) and the Center for Computing in Engineering and
Sciences at Unicamp (FAPESP grant \#2013/08293-7).

\section*{Funding}
This work was supported by the Pukyong National University Research
Fund in 202516520001. AFF acknowledges support from the Brazilian
Agency CNPq-Brazil (Grant number 302009/2025-6); S\~{a}o Paulo
Research Foundation (FAPESP) (Grant number \#2024/14403-4); and
Funda\c{c}\~{a}o de Apoio ao Ensino, Pesquisa e Extens\~{a}o –
FAEPEX/UNICAMP (Grant number \#3423/25).

\section*{Author Contributions}
S.J. and A.F.F. conceived the idea.
S.J. and A.F.F. performed the computational simulations and calculations. 
S.J., J.-W.H. and A.F.F. analyzed the results. 
S.J. wrote the original draft and S.J., J.-W.H. and A.F.F reviewed and edited the draft. 
S.J. and A.F.F supervised the work and acquired funding. 
All authors read and approved the final manuscript. 

\section*{Conflict of Interest Statement}

On behalf of all authors, the corresponding author states that there
is no conflict of interest.

\section*{Data Availability Statement}

Data available on reasonable request from the authors.

\bibstyle{sn-basic} 
\bibliography{main}

\end{document}